# Magnetic field induced transition from a vortex liquid to Bose metal in ultrathin a-MoGe thin film

Surajit Dutta, John Jesudasan and Pratap Raychaudhuri[*]

*Tata Institute of Fundamental Research, Homi Bhabha Road, Mumbai 400005, India.*

We identify a magnetic field induced transition from a vortex liquid to Bose metal in a 2-dimensional amorphous superconductor, *a*-MoGe, using a combination of magnetotransport and scanning tunnelling spectroscopy (STS). Below the superconducting transition, $T_c \sim 1.36$ K, the magnetoresistance isotherms cross at a nearly temperature independent magnetic field, $H_c^* \sim 36$ kOe. Above this field, the temperature coefficient of resistance is weakly negative, but the resistance remains finite as $T \rightarrow 0$, as expected in a bad metal. From STS conductance maps at 450 mK we observe a very disordered vortex lattice at very low fields that melts into a vortex liquid above 3 kOe. Up to $H_c^*$ the tunnelling spectra display superconducting gap and coherence peak over a broad background caused by electron-electron interactions, as expected in a vortex liquid. However, above $H_c^*$ the tunnelling spectra continue to display the gap but the coherence peak gets completely suppressed, suggesting that Cooper pairs lose their phase coherence. We conclude that $H_c^*$ demarcates a transition from a vortex liquid to Bose metal, that eventually transforms to a regular metal at a higher field $H^*$ where the gap vanishes in the electronic spectrum.

---

[*] E-mail: pratap@tifr.res.in

The survival of Cooper pairing even after the global phase coherence is destroyed has been a recurring theme in strongly disordered and 2-dimensional superconductors[1,2,3,4]. In zero field, the zero resistance state can get destroyed due to phase-fluctuations at the superconducting transition temperature, $T_c$, even though Cooper pairs continue to survive up to a much higher temperature[5,6,7], $T^*$, giving rise to the so called pseudogap state[8,9,10,11,12] between $T_c$ and $T^*$. On the other hand, when a magnetic field is applied at low temperatures, one observes another intriguing phenomenon, the superconductor (S) to bad-metal/insulator (BM/I) transition[13,14,15,16,17,18]. At a characteristic magnetic field, $H_c^*$, the magnetoresistance isotherms intersect, giving rise to a state with negative temperature coefficient of resistance ($R$) above this field. In very strongly disordered superconductors such as $InO_x$ and TiN this results in a strongly insulating state with diverging electrical resistance, sometimes called a superinsulator, which is believed to be the conjugate of the superconductor[19,20,21], where Cooper pairs instead of vortices are localised. On the other hand, when the disorder is more moderate, in several systems such as MoGe, Ta and NbN the resistance increases weakly with decreasing temperature and remains finite even as $T \to 0$; this state is more appropriately classified as a bad metal[22,23,24,14,25,26]. Whether Cooper pairs survive in this bad metal is an open question. Normally, a metallic state consisting of Cooper pairs is not expected since Cooper pairs are either in eigenstate of phase which gives a superconductor, or in eigenstate of number which gives an insulator. However, in recent years there have been suggestions that Cooper pairs could exist in a dissipative state called Bose metal[27,28,29,30,31,32,33,34,35] although its existence remains hotly debated[36,37,38]. Therefore, the metallic state above $H_c^*$, in such systems deserves careful attention.

In this letter, we investigate the field induced S-BM transition in a 2 nm thick amorphous $Mo_{70}Ge_{30}$ ($a$-MoGe) thin film, using a combination of low temperature scanning tunnelling spectroscopy (STS) and magneto-transport measurements. The primary advantage of STS is that it provides direct information on the local density of states. The central finding of this work is that the field induced bad metal state shows clear spectroscopic signatures of the existence of phase incoherent Cooper pairs as predicted for a Bose metal.

The *a*-MoGe film was grown by pulsed laser deposition. Details of sample growth and characterisation have been reported elsewhere[39,40,12]. STS measurements were performed using a home-built low-temperature scanning tunnelling microscope[41] (STM) operating down to 450 mK and up to a magnetic field of 90 kOe using a Pt-Ir normal metal tip. The tunnelling conductance, $G(V) = \frac{dI}{dV}\big|_V$, as a function of voltage ($V$), was measured using standard lock-in technique. To image vortices, spatially resolved $G(V)$ maps were recorded at a fixed voltage close to the coherence peak, where each vortex appears as a local minimum. We used an ultra-high-vacuum suitcase to transport the sample after deposition and transfer in the STM without exposure to air. After completing STS measurements, the sample was transported back to the deposition chamber in the same way and covered with a 1 nm thick Si protective layer before transport measurements. Magneto-transport measurements were performed in a conventional $^3$He cryostat. Since the magneto-transport properties of *a*-MoGe films are extremely susceptible to external electromagnetic radiation[42], all electrical feedthroughs leading to the sample were fitted with *RC* filters with a very low cut-off frequency of 100 Hz.

Fig. 1(a) shows the normalised conductance, $G_N(V) = \frac{G(V)}{G(4\ mV)}$, in zero field, averaged over $32 \times 32$ grid over $150\ nm \times 150\ nm$ area at various temperatures. Superconductivity manifests as a suppression of $G_N(V)$ for voltage, $|V| < \Delta/e$, and the appearance of coherence peaks at the gap edge. In disordered superconductors, we observe an additional V-shaped background that extends up to high bias[43,44,45], which originates from *e-e* Coulomb interactions described within Altshuler-Aronov (AA) theory[46,47,48]. In order to fit the data, incorporating both the AA correction and the effect of superconductivity, we adopt a procedure recently developed by Zemlicka et al.[49] At temperature *T*, the tunnelling conductance is given in terms of the single particle density of states of the sample, *N(E)*, by the relation[50],

$$G(V,T) \propto \int_{-\infty}^{\infty} dE \frac{1}{k_B T} \frac{e^{\frac{E+eV}{k_B T}}}{\left(1+e^{\frac{E+eV}{k_B T}}\right)^2} N(E) \ , \qquad (1)$$

where $k_B$ is the Boltzmann constant. Here we assume the density of states of the Pt-Ir tip to be energy independent, which is a good approximation within 10s of meV from the Fermi level. In a normal metal, AA corrections modify the bare density of states, $N_N$ as[49],

$$N_N^{AA}(E) = N_N\big(1 + \tilde{\lambda}_0[f_d(E,\Gamma_0) - \lambda_r\{f_d(E,\Gamma_1) + f_d(E+E_z,\Gamma_1) + f_d(E-E_z,\Gamma_1)\}]\big) \quad (2a)$$

$$f_d(E,\Gamma_n) = -\frac{1}{2}\int_0^{\frac{\Gamma}{k_BT}} dx \frac{x}{x^2+\left(\frac{\Gamma_n}{k_BT}\right)^2} \frac{Sinh(x)}{Cosh(x)+Cosh\left(\frac{E}{k_BT}\right)} \quad (2b)$$

where $\Gamma = \hbar/\tau$, $\tau$ is the transport scattering time, $E_z = 2\mu_B H$ ($\mu_B$ is Bohr magneton) is the Zeeman energy when there is an applied magnetic field $H$, and $\Gamma_0$ and $\Gamma_1$ account for the broadening due to energy and spin scattering. We use $\Gamma_0$, $\Gamma_1$, $\tilde{\lambda}_0$ and $\lambda_r$ as adjustable parameters. In Fig. 1(a) we can fit the data for $T \geq 3.7\ K$ using eqn. 1 with $N(E) = N_N^{AA}(E)$; we take $E_z = 0$, $\Gamma_0 = \Gamma_1 = 0.2\ meV$. It is important to note that the parameters related to AA fitting do not vary more than ±10% from their mean value over the entire range of temperature and magnetic field which arises from statistical error. The complete set of best fit parameters is given in the Supplementary Material[51]. In principle, $\Gamma$ is also a fitting parameter which we take as $\Gamma \sim 20\ eV$ based on $\tau \sim 3 \times 10^{-17} s$ estimated from resistivity. However, for such a large value of $\Gamma (\gg \{k_BT,\Gamma_0,\Gamma_1\})$, eqn. (2b) is insensitive to its precise value and practically indistinguishable from setting the integration limit to infinity. For $T < 3.7\ K$, we cannot fit the data with AA contributions alone and need to incorporate the effect of superconductivity. In BCS theory where the density of state in the normal metal is assumed to be energy independent, the single particle density of states is given by[50],

$N_S^{BCS}(E) = \left|Re\left\{\frac{E-i\Gamma^D}{\sqrt{(E-i\Gamma^D)^2-\Delta^2}}\right\}\right|$, where $\Gamma^D$ is a phenomenological broadening parameter[52] that accounts for non-thermal sources of broadening. When this correction is applied on the top of AA corrections, the single particle density of states is given by[49],

$$N(E) = N_S^{BCS}(E) N_N^{AA}(\Omega), \quad (3)$$

where $\Omega = Re\left[\sqrt{(E - i\Gamma^D)^2 - \Delta^2}\right]$; $\Omega$ in the second term ensures conservation in the number of states (and charge) above and below the superconducting transition. Using this form for $N(E)$ and $\Delta$, $\Gamma^D$ as fitting parameters, we can fit the $G_N(V)$ spectra down to 450 mK keeping $\Gamma_0$ and $\Gamma_1$ same as the normal state value and (Fig. 1(a)). Figure 1(b) shows the temperature variation of $\Delta$ and $\Gamma^D$ along with the temperature variation of the sheet resistance, $R_s$. $T_c \sim 1.36$ K is defined as the temperature where $R_s$ is 0.05% of its normal state value is much lower than $T^* \sim 3.7$ K where $\Delta$ vanishes, showing the existence of a large pseudogap state. The zero bias conductance map (inset, Fig. 1(a)) shows large variations, forming puddle like structures consistent with earlier reports[12].

We now turn our attention to the effect of magnetic field. Figure 2(a) shows $R_s$ as a function of $H$. Below $T_c$, all the $R_s$-$H$ curves cross close to $H_c^* \sim 36\ kOe$ which demarcates the S-BM transition. Concentrating first at low temperatures and low fields, we observe finite $R_s$ above 3 kOe (Fig. 2(b)) at 450 mK. We confirm that this $R_s$ indeed corresponds to the linear resistance, by plotting in the same graph $R_{lin} = \frac{dI}{dV}\bigg|_{V \to 0}$ calculated from current-voltage (*I-V*) characteristic measurements (*inset* Fig. 2(b)). The appearance of a finite linear resistance at subcritical current indicates the formation of vortex liquid (VL) state[53]. Further evidence of VL is obtained from STS images of the vortex state (Fig. 2(d)). At 1 kOe we can resolve individual vortices even though the spatial configuration is extremely disordered; at 5 kOe some vortices are still visible but at many places we also observe blurred or elongated structures indicating considerable motion of vortices during the acquisition time of the image; at 10 kOe the vortex system is deep into VL and we cannot resolve individual vortices anymore. From $R_s$ vs. $T$ at different $H$ (Fig. 2(c)) we observe that for $H > H_c^*$, $\frac{dR_s}{dT} < 0$ down to 300 mK, even though the sheet conductance $G_s\left(=\frac{1}{R_s}\right)$ extrapolates to a finite value as $T \to 0$, characteristic of a bad metal.

To understand the evolution of the superconducting state at higher fields, we investigate the $G_N(V)\ vs. V$ spectra, spatially averaged on a $32 \times 32$ grid over $150\ nm \times 150\ nm$ area (Fig. 3(a)). For $H > 70\ kOe$ the tunnelling spectra can be fitted using $N(E) = N_N^{AA}(E)$. We define this field as $H^*$. Below $H^*$, we observe two distinct magnetic field regimes (Fig. 3(b)). For $H < H_c^*$ we can fit the tunnelling spectra using $N(E)$ in eqn. 3 using $\Delta$ and $\Gamma^D$ shown in the *inset* of Fig. 3(d). For $H^* \geq H > H_c^*$ this form is no longer adequate; we observe that the fitted value is larger in the voltage range corresponding to the energy where the coherence peaks are expected in $N_s^{BCS}(E)$. This deviation results from a complete suppression of the superconducting coherence peaks above $H_c^*$. To illustrate this point we modify $N_s^{BCS}(E)$ in the following way: We use the empirical form, $N_s^{Mod}(E) = \alpha_0 - A\exp(-BE^2)$, and adjust $\alpha_0$, $A$ and $B$ such that $N_s^{Mod}(E) \approx N_s^{BCS}(E)$ when $N_s^{Mod}(E) \lesssim 0.95$, but asymptotically approaches 1 instead of exhibiting the coherence peak at higher values (Fig. 3(c)). Using $N(E) = N_s^{Mod}(E)N_N^{AA}(\Omega)$ results in an excellent fit of the tunneling curves above $H_c^*$ (Fig. 3(b)). In Fig. 3(d), we plot the coherence peak height (CPH), defined as $max\{N_s^{BCS/Mod}(E)\}$-1. With increasing field CPH gradually reduces and becomes zero above $H_c^*$.

To understand the physical implication of these results we note that, while the superconducting energy gap is a direct manifestation of the pairing of electrons forming Cooper pairs, the superconducting state requires another ingredient, namely, the global phase coherence among Cooper pairs, which is related to the appearance of the coherence peaks at the gap edge[50]. If Cooper pairs lack phase coherence, it is expected that the single particle spectrum will remain gapped, but the coherence peaks will get suppressed[2,4,7,54]. In magnetic field, the nucleation of vortices complicates this scenario. Here both the gap and the coherence peaks vanish inside the vortex core but appear as one goes further from the vortex centre. In a VL, where vortices are rapidly moving, a slow measurement like STS measures the time averaged tunnelling spectrum which has contribution from both inside and outside the core. This broadens the spectrum by partially filling the gap and suppressing the coherence peaks. It has been shown[55] that the average spectrum in the mixed state can be captured by adjusting $\Gamma^D$. This accounts for the gradual increase in $\Gamma^D$ and suppression of CPH below $H_c^*$. On the other hand, the abrupt disappearance of the CPH for $H^* \geq H > H_c^*$, with no commensurate filling of the gap indicates that the Cooper pairs no longer have phase coherence. In conjunction with $G_s$ in the $T \rightarrow 0$ limit, this implies the formation of a dissipative state made of Cooper pairs, i.e. a Bose metal. We would like to note that this state is distinct from the zero field pseudogap state, where, as shown before, the spatially averaged spectra do not undergo any qualitative change across $T_c$. The origin of the pseudogap state lies in the emergence of tens of nanometer sized superconducting puddles[56,57] separated by insulating regions and thermal fluctuations between these puddles that destroy the zero resistance state[12,58]. Here, even though we observe inhomogeneity in $G_N(0)$, the complete suppression of CPH in the average spectrum suggests that the coherence peak is uniformly suppressed everywhere. This has been further confirmed[51] by selectively fitting spectra in regions of high and low $G_N(0)$.

In Figure 4 we show the phase boundaries of $H_c^*$ and $H^*$ in the $H$-$T$ parameter space. The $H^*$ boundary is determined from STS measurement and is defined as the highest magnetic field (at fixed temperature) or highest temperature (at fixed field) where the tunnelling spectra cannot be fitted with the AA contribution alone. The $H_c^*$ boundary on the other hand is defined from the magnetic field at which two magnetoresistance curves at successive temperatures intersect. As expected for a quantum phase transition, this boundary is nearly temperature independent till about $T_c$; above $T_c$ it closely follows $H^*$ boundary. Finally, we define the vortex lattice melting boundary, $H^m$, above which a finite linear resistance appears. The temperature variation of $H^m$ is described very well by the formula for thermal melting[59,51] of the vortex lattice if one substitutes $H_c^*$ for the upper critical field.

To place things into perspective all our results lead to the conclusion that above and below $H_c^*$ there exist two distinct dissipative states demarcated by distinct transport and spectroscopic signatures. While the state below $H_c^*$ is a classical VL, the question remains as to what the other state which we phenomenologically classify as a Bose metal is. Though we cannot conclusively settle this question here, one possibility is that it is a quantum VL. In a recent paper[60], it has been shown that vortices in a 20 nm thick *a*-MoGe film undergo quantum zero-point fluctuations, whose fractional amplitude with respect to intervortex separation increases with increasing magnetic field. Here, a similar analysis shows[51] that the zero point fluctuation amplitude of the vortices is expected to be about half the inter-vortex separation at $H_c^*$, such that quantum tunneling could be the dominant mechanism of vortex motion above this field.

Whether such a state would display the complete suppression of the coherence peak needs to be theoretically investigated. The other point to note is that in our experiment the Bose metal exists only for $T \lesssim T_c$, when the zero field state can sustain a finite supercurrent, even though fragmented superconducting puddles continue to exist up to $T^*$.

In summary, we have shown spectroscopic evidence of the existence of a magnetic field induced Bose metal in ultrathin *a*-MoGe thin film. These results have relevance to other systems like High Temperature superconductors[30] where Bose metal states have been reported. We hope that our results will bolster further theoretical investigations on how such a state emerges in a 2-dimensional superconductor.

Acknowledgements: We thank Nandini Trivedi for valuable suggestions. This paper was supported by the Department of Atomic Energy, Government of India (Grant No. 12-R&D-TFR-5.10-0100).

[1] A. Ghosal, M. Randeria, and N. Trivedi, *Role of Spatial Amplitude Fluctuations in Highly Disordered s-wave Superconductors*, Phys. Rev. Lett. **81**, 3940 (1998).

[2] M. V. Feigel'man, L. B. Ioffe, V. E. Kravtsov, and E. A. Yuzbashyan, *Eigenfunction Fractality and Pseudogap State near the Superconductor-Insulator Transition*, Phys. Rev. Lett. **98**, 027001 (2007).

[3] Y. Dubi, Y. Meir and Y. Avishai, *Nature of the superconductor–insulator transition in disordered superconductors*, Nature **449**, 876 (2007).

[4] M. V. Feigel'man, L. B. Ioffe, V. E. Kravtsov, and E. Cuevas, *Fractal superconductivity near localization threshold,* Ann. Phys. **325**, 1390 (2010).

[5] V. J. Emery and S. A. Kivelson, *Importance of phase fluctuations in superconductors with small superfluid density*, Nature **374**, 434 (1995).

[6] A. Ghosal, M. Randeria, and N. Trivedi, *Inhomogeneous pairing in highly disordered s-wave superconductors*, Phys. Rev. B **65**, 014501 (2001).

[7] K. Bouadim, Y. L. Loh, M. Randeria and N. Trivedi, *Single- and two-particle energy gaps across the disorder-driven superconductor–insulator transition*, Nat. Phys. **7**, 884 (2011).

[8] B. Sacépé, C. Chapelier, T. I. Baturina, V. M. Vinokur, M. R. Baklanov and M. Sanquer, *Pseudogap in a thin film of a conventional superconductor,* Nat. Commun. **1**, 140 (2010).

[9] M. Mondal, A. Kamlapure, M. Chand, G. Saraswat, S. Kumar, J. Jesudasan, L. Benfatto, V. Tripathi and P. Raychaudhuri, *Phase fluctuations in a strongly disordered s-wave NbN superconductor close to the metal-insulator transition*, Phys. Rev. Lett. **106**, 047001 (2011).

[10] B. Sacepe, T. Dubouchet, C. Chapelier, M. Sanquer, M. Ovadia, D. Shahar, M. Feigel'man and L. Ioffe, *Localization of preformed Cooper pairs in disordered superconductors*, Nat. Phys. **7**, 239 (2011).

[11] T. Dubouchet, B. Sacépé, J. Seidemann, D. Shahar, M. Sanquer and C. Chapelier, *Collective energy gap of preformed Cooper pairs in disordered superconductors*, Nat. Phys. **15**, 233 (2019).

[12] S. Mandal, S. Dutta, S. Basistha, I. Roy, J. Jesudasan, V. Bagwe, L. Benfatto, A. Thamizhavel, and P. Raychaudhuri, *Destruction of superconductivity through phase fluctuations in ultrathin a-MoGe films*, Phys. Rev. B **102**, 060501(R) (2020).

[13] A. F. Hebard and M. A. Paalanenn, *Magnetic-Field-Tuned Superconductor-Insulator Transition in Two-Dimensional Films*, Phys. Rev. Lett. **65,** 927 (1990).

[14] A. Yazdani and A. Kapitulnik, *Superconducting-Insulating Transition in Two-Dimensional a-MoGe Thin Films*, Phys. Rev. Lett. **74**, 3037 (1995).

[15] E. Bielejec and W. Wu, *Field-Tuned Superconductor-Insulator Transition with and without Current Bias*, Phys. Rev. Lett. **88**, 206802 (2002).

[16] V. F. Gantmakher, M. V. Golubkov, V. T. Dolgopolov, A. A. Shashkin, and G. E. Tsydynzhapov, *Observation of the Parallel-Magnetic-Field-Induced Superconductor–Insulator Transition in Thin Amorphous $InO_x$ Films*, JETP Lett. **71**, 160 (2000).

[17] G. Sambandamurthy, L. W. Engel, A. Johansson, and D. Shahar, *Superconductivity-Related Insulating Behavior,* Phys. Rev. Lett. **92**, 107005 (2004).

[18] T. I. Baturina, D. R. Islamov, J. Bentner, C. Strunk, M. R. Baklanov and A. Satta, *Superconductivity on the localization threshold and magnetic-field-tuned superconductor-insulator transition in TiN films*, JETP Lett. **79**, 337 (2004).

[19] V. M. Vinokur, T. I. Baturina, M. V. Fistul, A. Yu. Mironov, M. R. Baklanov and C. Strunk, *Superinsulator and quantum synchronization*, Nature **452**, 613 (2008).

[20] M. Ovadia, D. Kalok, B. Sacépé and D. Shahar, *Duality symmetry and its breakdown in the vicinity of the superconductor–insulator transition*, Nat. Phys. **9**, 415 (2013).

[21] S. Sankar, V. M. Vinokur, and V. Tripathi, *Disordered Berezinskii-Kosterlitz-Thouless transition and superinsulation*, Phys. Rev. B **97**, 020507(R) (2018).

[22] N. Mason and A. Kapitulnik, *Dissipation Effects on the Superconductor-Insulator Transition in 2D Superconductors*, Phys. Rev. Lett. **82,** 5341 (1999).

[23] M. Chand, G. Saraswat, A. Kamlapure, M. Mondal, S. Kumar, J. Jesudasan, V. Bagwe, L. Benfatto, V. Tripathi and P. Raychaudhuri, *Phase diagram of the strongly disordered s-wave superconductor NbN close to the metal-insulator transition*, Phys. Rev. B **85**, 014508 (2012).

[24] Yongguang Qin, Carlos L. Vicente, and Jongsoo Yoon, *Magnetically induced metallic phase in superconducting tantalum films*, Phys. Rev. B **73**, 100505(R) (2006).

[25] I. Zaytseva, A. Abaloszew, B. C. Camargo, Y. Syryanyy and M. Z. Cieplak, *Upper critical field and superconductor-metal transition in ultrathin niobium films*, Scientific Reports **10**, 19062 (2020).

[26] W. Liu, L. Pan, J. Wen, M. Kim, G. Sambandamurthy, and N. P. Armitage, *Microwave Spectroscopy Evidence of Superconducting Pairing in the Magnetic-Field-Induced Metallic State of $InO_x$ Films at Zero Temperature*, Phys. Rev. Lett. **111**, 067003 (2013).

[27] P. Phillips, D. Dalidovich, *The elusive Bose metal*, Science **302**, 243 (2003).

[28] J. Wu and P. Phillips, *Vortex glass is a metal: Unified theory of the magnetic-field and disorder-tuned Bose metals,* Phys. Rev. B **73**, 214507 (2006).

[29] B. Spivak, P. Oreto, and S. A. Kivelson, *Theory of quantum metal to superconductor transitions in highly conducting systems*, Phys. Rev. B **77**, 214523 (2008).

[30] C. Yang, Y. Liu, Y. Wang, L. Feng, Q. He, J. Sun, Y. Tang, C. Wu, J. Xiong, W. Zhang, X. Lin, H. Yao, H. Liu, G. Fernandez, J. Xu, J. M. Valles Jr., J. Wang and Y. Li, *Intermediate bosonic metallic state in the superconductor-insulator transition*, Science **366**, 1505 (2019).

[31] T. Ren and A. M Tsvelik, *How magnetic field can transform a superconductor into a Bose metal*, New J. Phys. **22**, 103021 (2020).

[32] P. Phillips and D. Dalidovich, *Short-range interactions and a Bose metal phase in two dimensions*, Phys. Rev. B **65**, 081101(R) (2002).

[33] S. Doniach and D. Das, *The Bose metal - A Commentary*, Braz. J. Phys. **33**, 740 (2003).

[34] M.C. Diamantini, A.Yu. Mironov, S.M. Postolova, X. Liu, Z. Hao, D.M. Silevitch, Ya. Kopelevich, P. Kim, C.A. Trugenberger, V.M. Vinokur, *Bosonic topological insulator intermediate state in the superconductor-insulator transition*, Phys. Lett. A **384,** 126570 (2020).

[35] D. Das and S. Doniach, *Bose metal: Gauge-field fluctuations and scaling for field-tuned quantum phase transitions*, Phys. Rev. B **64**, 134511 (2001).

[36] I. Tamir, A. Benyamini, E. J. Telford, F. Gorniaczyk, A. Doron, T. Levinson, D. Wang, F. Gay, B. Sacépé, J. Hone, K. Watanabe, T. Taniguchi, C. R. Dean, A. N. Pasupathy, D. Shahar, *Sensitivity of the superconducting state in thin films*, Sci Adv. 5, eaau3826 (2019).

[37] P. W. Phillips, *Free at last: Bose metal uncaged*, Science **366,** 1450 (2019).

[38] X. Zhang, B. Hen, A. Palevski and A. Kapitulnik, *Robust anomalous metallic states and vestiges of self-duality in two-dimensional granular In-InO$_x$ composites*, npj Quantum Materials **6**, 30 (2021).

[39] I. Roy, S. Dutta, A. N. Roy Choudhury, S. Basistha, I. Maccari, S. Mandal, J. Jesudasan, V. Bagwe, C. Castellani, L. Benfatto, and P. Raychaudhuri, *Melting of the Vortex Lattice through Intermediate Hexatic Fluid in an a-MoGe Thin Film***,** Phys. Rev. Lett. **122**, 047001 (2019).

[40] S. Dutta, I. Roy, S. Basistha, S. Mandal, J. Jesudasan, V. Bagwe and P. Raychaudhuri, *Collective flux pinning in hexatic vortex fluid in a-MoGe thin film*, J. Phys.: Condens. Matter **32**, 075601 (2020).

[41] A. Kamlapure, G. Saraswat, S. C. Ganguli, V. Bagwe, P. Raychaudhuri, and S. P. Pai, *A 350 mK, 9 T scanning tunneling microscope for the study of superconducting thin films on insulating substrates and single crystals*, Rev. Sci. Instrum. **84**, 123905 (2013).

[42] S. Dutta, I. Roy, S. Mandal, J. Jesudasan, V. Bagwe, and P. Raychaudhuri, *Extreme sensitivity of the vortex state in a-MoGe films to radio-frequency electromagnetic perturbation*, Phys. Rev. B **100**, 214518 (2019).

[43] S. P. Chockalingam, M. Chand, A. Kamlapure, J. Jesudasan, A. Mishra, V. Tripathi and P. Raychaudhuri, *Tunneling studies in a homogeneously disordered s-wave superconductor: NbN,* Phys. Rev. B **79**, 094509 (2009).

[44] C. Carbillet, V. Cherkez, M. A. Skvortsov, M. V. Feigel'man, F. Debontridder, L. B. Ioffe, V. S. Stolyarov, K. Ilin, M. Siegel, D. Roditchev, T. Cren, and C. Brun, *Spectroscopic evidence for strong correlations between local resistance and superconducting gap in ultrathin NbN films*, Phys. Rev. B **102**, 024504 (2020).

[45] S. V. Postolova, A. Yu. Mironov, V. Barrena, J. Benito-Llorens, J. G. Rodrigo, H. Suderow, M. R. Baklanov, T. I. Baturina, and V. M. Vinokur, *Superconductivity in a disordered metal with Coulomb interactions*, Phys. Rev. Research **2**, 033307 (2020).

[46] B. L. Altshuler and A. G. Aronov, in Electron-Electron Interactions in Disordered Systems, Modern Problems in Condensed Matter Sciences edited by A. Efros and M. Pollak (Elsevier, 1985), pp. 1–153.

[47] B. L. Altshuler, A. G. Aronov, and P. A. Lee, *Interaction Effects in Disordered Fermi Systems in Two Dimensions*, Phys. Rev. Lett. **44**, 1288 (1980).

[48] B.L.Altshuler and A.G.Aronov, *Zero Bias Anomaly in Tunnel Resistance and Electron-electron Interaction*, Solid State Commun. **36**, 115 (1979).

[49] M. Žemlička, M. Kopčík, P. Szabó, T. Samuely, J. Kačmarčík, P. Neilinger, M. Grajcar, and P. Samuely, *Zeeman-driven superconductor-insulator transition in strongly disordered MoC films: Scanning tunneling microscopy and transport studies in a transverse magnetic field*, Phys. Rev. B **102**, 180508(R) (2020).

[50] M. Tinkham, Introduction to Superconductivity (McGrawHill, Singapore, 1996).

[51] See supplementary material for (i) Fit of the spectra at 450 mK, 40 kOe and regions of high and low $G_N(0)$ and (ii) fit of the $H^m$ line in the $H$-$T$ parameter space with the theory of thermal vortex melting, (iii) the complete table of best fit parameters for the tunneling spectra, (iv) analysis of possible zero point fluctuation of vortices and (v) scaling analysis of the Magnetoresistance curves.

[52] R. C. Dynes, V. Narayanamurti, and J. P. Garno, *Direct Measurement of Quasiparticle-Lifetime Broadening in a Strong-Coupled Superconductor*, Phys. Rev. Lett. **41**, 1509 (1978).

[53] V. M. Vinokur, M. V. Feigel'man, V. B. Geshkenbein, and A. I. Larkin, *Resistivity of High-Tc Superconductors in a Vortex-Liquid State*, Phys. Rev. Lett. **65**, 259 (1990).

[54] G. Lemarie, A. Kamlapure, D. Bucheli, L. Benfatto, J. Lorenzana, G. Seibold, S. C. Ganguli, P. Raychaudhuri, and C. Castellani, *Universal scaling of the order-parameter distribution in strongly disordered superconductors,* Phys. Rev. B **87**, 184509 (2013).

[55] S. Mukhopadhyay, Goutam Sheet, P Raychaudhuri, *Magnetic-field dependence of superconducting energy gaps in $YNi_2B_2C$: Evidence of multiband superconductivity,* H Takeya, Phys. Rev. B **72**, 014545 (2005).

[56] B. Sacépé, C. Chapelier, T. I. Baturina, V. M. Vinokur, M. R. Baklanov, and M. Sanquer, *Disorder-Induced Inhomogeneities of the Superconducting State Close to the Superconductor-Insulator Transition*, Phys. Rev. Lett. **101**, 157006 (2008).

[57] A. Kamlapure, T. Das, S. C. Ganguli, J. B. Parmar, S. Bhattacharyya and P. Raychaudhuri, *Emergence of nanoscale inhomogeneity in the superconducting state of a homogeneously disordered conventional superconductor*, Sci. Rep. **3**, 2979 (2013).

[58] M. Mondal, A. Kamlapure, S. C. Ganguli, J. Jesudasan, V. Bagwe, L. Benfatto and P. Raychaudhuri, *Enhancement of the finite-frequency superfluid response in the pseudogap regime of strongly disordered superconducting films,* Sci. Rep. **3**, 1357 (2013).

[59] G. Blatter, M. V. Feigel'man, V. B. Geshkenbein, A. I. Larkin, and V. M. Vinokur, *Vortices in high-temperature superconductors*, Rev. Mod. Phys. **66**, 1125 (1994).

[60] S. Dutta, I. Roy, J. Jesudasan, S. Sachdev, and P. Raychaudhuri, *Evidence of zero-point fluctuation of vortices in a very weakly pinned a-MoGe thin film*, Phys. Rev. B **103**, 214512 (2021).

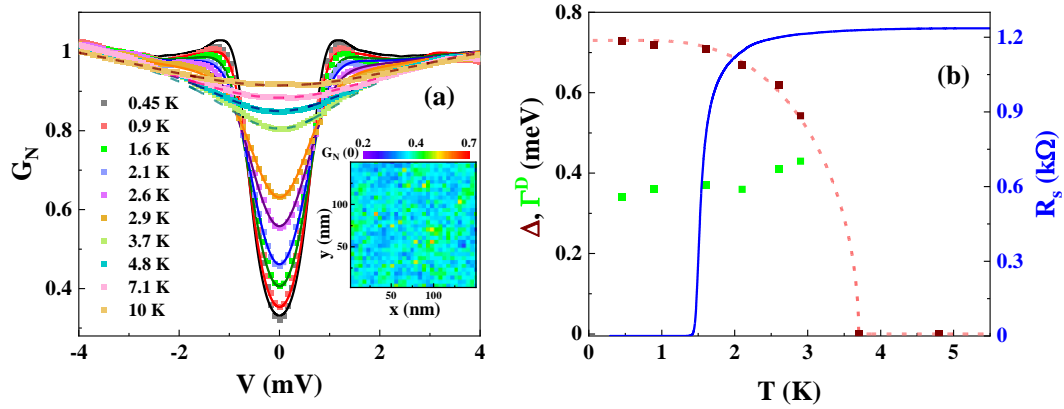

**Figure 1|** (a) $G_N(V)$-$V$ tunneling spectra in zero field at different temperatures along with theoretical fit using eqn. 1. At 3.7 K and above the spectra are fitted with AA corrections alone (dashed line) whereas at lower temperatures additional contribution from superconductivity has to be incorporated (solid lines). (*inset*) spatial map of $G_N(0)$ at 450mK. (b) Temperature dependence of superconducting energy gap $\Delta$ and $\Gamma^D$ (left axis) and sheet resistance $R_s$ (right axis) at zero magnetic field; the dashed line is a guide to the eye that mimics the BCS temperature variation of $\Delta$.

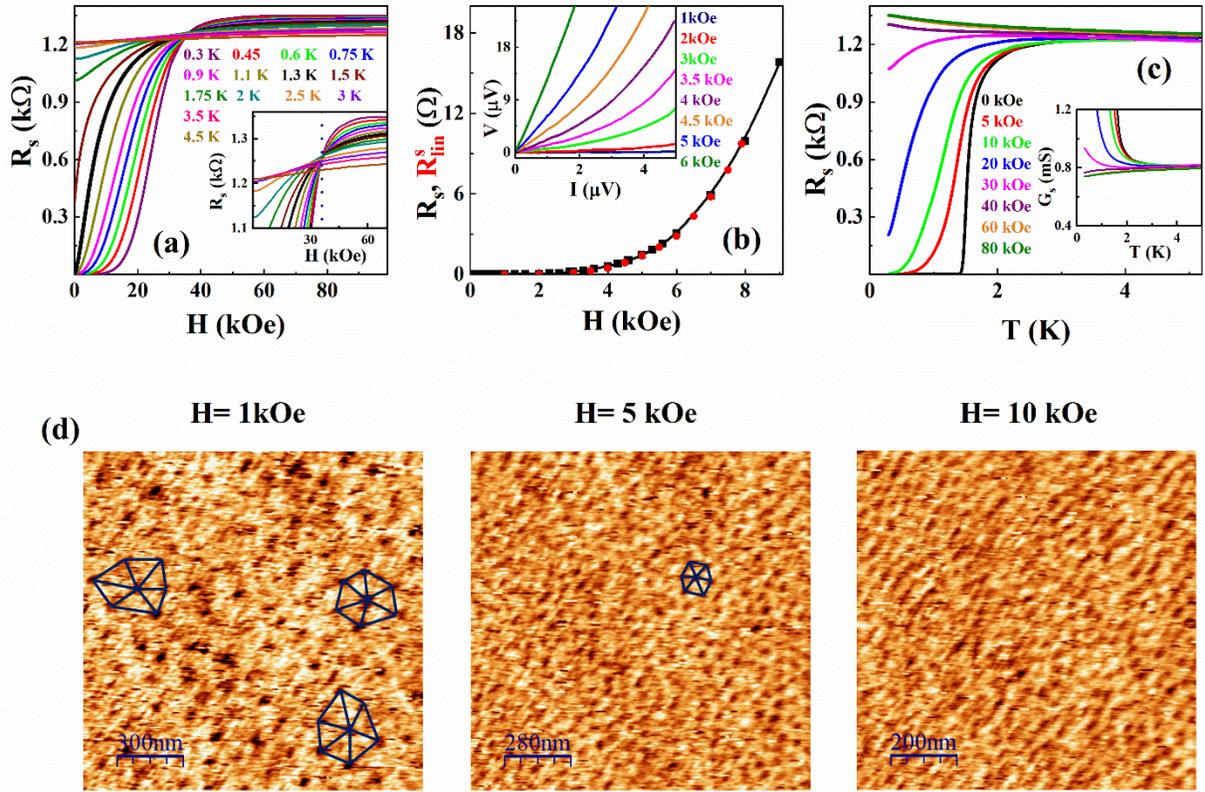

**Figure2|** (a) Magnetic field dependence of sheet resistance $R_s$ at different temperatures. (*inset*) expanded view of same plot close to $H_c^*$ (blue dotted line). (b) Plot of $R_s$ (black square) and $R_s^{lin}$ (red circle) vs $H$ at low magnetic fields at 450 mK; (*inset*) *I-V* characteristics at the same temperatures. (c) Temperature dependence of $R_s$ at various magnetic fields. (*inset*) shows corresponding temperature variation of sheet conductance $G_s$. (d) Spatial maps of $G_N(V)$ at the coherence peak voltage $V = 1.2$ mV at 450 mK for $H = 1$ kOe, 5 kOe, and 10 kOe. At 1 kOe we observe a disordered vortex lattice; at 5 kOe individual vortices can be resolved only at some locations but in the rest of the area individual vortices cannot be identified; at 10 kOe we can no longer resolve individual vortices indicating that the vortex lattice has completely molten. Representative hexagonal coordination is shown at few representative locations at 1 kOe and 5 kOe.

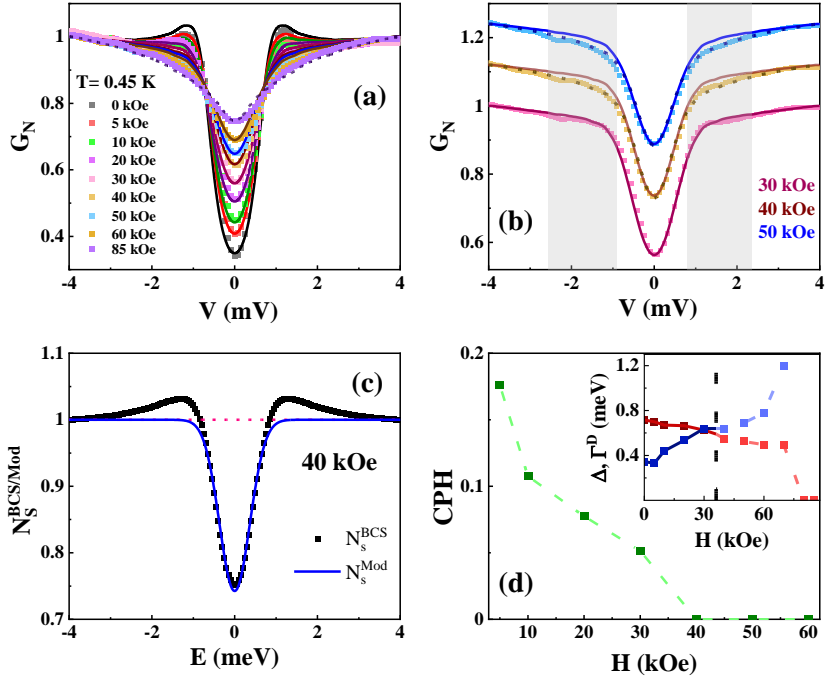

**Figure3|** (a) $G_N(V)$ vs $V$ in different magnetic fields at 450mK along with theoretical fit using eqn. 1. For $H > 70$ kOe the spectra are fitted with AA corrections alone (dashed line) whereas at lower temperatures additional contribution from superconductivity has to be incorporated (solid lines). (b) Fit of $G_N(V)$ vs $V$ at $H = 30$ kOe, 40 kOe and 50 kOe using $N_S^{BCS}(E)$ (solid line) and $N_S^{Mod}(E)$ (dashed line) respectively; successive spectra following 30 kOe are shifted upward by 0.12 for clarity. For 40 and 50 kOe ($H > H_c^*$) the fit using $N_S^{BCS}(E)$ deviates significantly in the voltage range where the coherence peak appears (shaded grey region). (c) Plot of $N_S^{BCS}(E)$ (Black dot) and $N_S^{Mod}(E)$ (blue solid line), used for the theoretical fit in (b) at 40 kOe. (d) Magnetic field dependence of coherence peak height (CPH). (*inset*) Variation of superconducting energy gap $\Delta$, $\Gamma^D$ as a function of magnetic field at 450 mK; for $H > H_c^*$, these values correspond to the values used in $N_S^{BCS}(E)$ from which $N_S^{Mod}(E)$ is constructed.

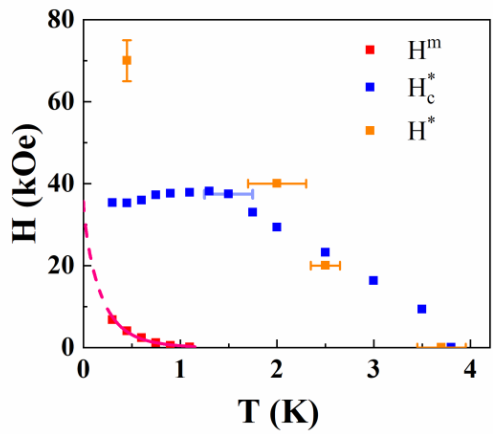

**Figure4|** Phase diagram showing the temperature evolution of $H^*$ (orange square), $H_c^*$ (blue square), and $H^m$ (red square); the dashed line is the fit to thermal melting line. The Bose metal phase is realized between $H_c^*$ and $H^*$. In between $H^m$ and $H_c^*$ we have a vortex liquid.

# Supplementary Material

**I. Spatial variation of tunneling spectra at H = 40 kOe**

In this section, we analyze the spatial variation of the tunneling spectra at 40 kOe which is smallest magnetic field at 450 mK at which the coherence peak in the superconducting density of states disappears. As it can be seen from Fig. S1(a) in this field $G_N(0)$ has considerable variation over 10-s of nanometer length scales. We have illustrated the fitting of the selective area tunneling spectra at H= 40 kOe and T= 450mK. We separately analyze the spectra averaged from two kinds of regions: (i) regions where $G_N(0) \geq 0.68$ and (ii) regions where $G_N(0) \leq 0.53$. We observe that both kind of spectra can be fitted perfectly using $N_s^{Mod}(E) = \alpha_0 - A exp(-BE^2)$ as described in the main text (Fig. S1 (b)). The spectrum averaged from region (i) can also be fitted almost equally well using $N_s^{BCS}(E)$ if we take $\Gamma^D$ to be roughly 2 times $\Delta$, but such large value of $\Gamma^D$ effectively implies that the coherence peak is completely suppressed. The best fit parameters for the two curves are given in Table S1.

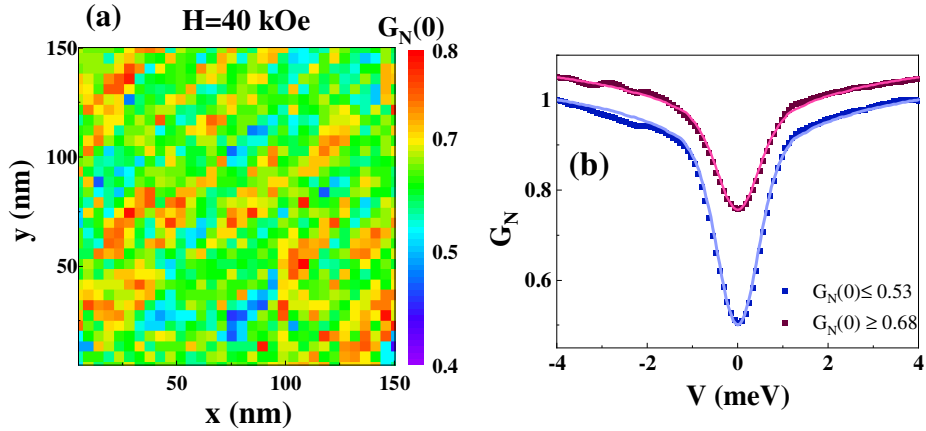

Fig. S1| (a) Spatial map of $G_N(0)$ at H= 40 kOe and T= 450 mK. (b) Corresponding $G_N$ -V tunneling spectra at averaged over regions of low $G_N(0)$ (blue dot) and high $G_N(0)$ (purple dot) along with theoretical fit using eqn. (1).

Table S1: Best fit parameters for the two curves shown in Fig. S1. $\Delta$ and $\Gamma^D$ are the starting values used in $N_S^{BCS}(E)$ from which $N_S^{Mod}(E)$ is constructed.

| H (kOe) | T (K) | ZBC ($G_N$ (0)) | $\Gamma$ (meV) | $\Gamma_0$ (meV) | $\tilde{\lambda}_0$ | $\lambda_r$ | $\Delta$ (meV) | $\Gamma^D$ (meV) |
|---|---|---|---|---|---|---|---|---|
| 40 | 0.45 | $G_N(0) \leq 0.53$ | $10^4$ | 0.2 | 0.072 | -0.1 | 0.71 | 0.59 |
| | | $G_N(0) \geq 0.68$ | $10^4$ | 0.2 | 0.066 | -0.1 | 0.52 | 0.92 |

## II. Thermal melting of vortex matter

In fig 4 (main manuscript), we mentioned that the temperature dependence of $H^m$ follows the melting transition line driven by thermal fluctuations. The melting line is theoretically obtained by solving the given equation[1],

$$\frac{\sqrt{b_m}}{1-b_m}\frac{t}{\sqrt{1-t}}\left[\frac{4(\sqrt{2}-1)}{\sqrt{1-b_m}}+1\right] = 2\pi\frac{C_L^2}{\sqrt{Gi}}$$

Where, the reduced melting field, $b_m = \frac{H_m}{H_{c2}}$, $t = \frac{T}{T_c}$ is the reduced temperature, $C_L$ is the Lindemann constant and $Gi$ is the Ginzburg-Lenvanyuk number. While fitting, $\frac{C_L^2}{\sqrt{Gi}}$ has been considered as the fitting parameter and the obtained melting line coincides well with the experimental data for $\frac{C_L^2}{\sqrt{Gi}} = 0.065$ (shown in fig. 4). $Gi$ is calculated using the given relation[2] (for two-dimensional disordered system),

$$Gi = \frac{e^2}{23\hbar}R_S^N = 0.013$$

$R_S^N (= R_S^{8K})$ is the normal state sheet resistance. Using the value of $Gi$, we find the Lindemann constant, $C_L \cong 0.09$ which is close to the expected value.

## III. Complete list of Best fit parameters for Tunneling spectra

| H (kOe) | T (K) | $\Gamma$ (meV) | $\Gamma_0$ (meV) | $\tilde{\lambda}_0$ | $\lambda_r$ | $\Delta$ (meV) | $\Gamma^D$ (meV) | Form of $N_S$ (E) |
|---|---|---|---|---|---|---|---|---|
| | 0.45 | | | 0.0765 | -0.102 | 0.73 | 0.34 | $N_S^{BCS}(E)$ |
| | 0.9 | | | 0.0765 | -0.102 | 0.72 | 0.36 | |

| H (kOe) | T (K) | $\Gamma$ (meV) | $\Gamma_0$ (meV) | $\tilde{\lambda}_0$ | $\lambda_r$ | $\Delta$ (meV) | $\Gamma^D$ (meV) | Form of $N_S(E)$ |
|---|---|---|---|---|---|---|---|---|
| 0 | 1.6 | $10^4$ | 0.2 | 0.0765 | -0.102 | 0.71 | 0.37 | |
| | 2.1 | | | 0.0765 | -0.102 | 0.67 | 0.36 | |
| | 2.6 | | | 0.0765 | -0.102 | 0.62 | 0.41 | |
| | 2.9 | | | 0.0765 | -0.102 | 0.543 | 0.43 | |
| | 3.7 | | | 0.083 | -0.116 | | | Fitted with AA correction only |
| | 4.8 | | | 0.078 | -0.107 | | | |
| | 5.8 | | | 0.078 | -0.107 | | | |
| | 7.1 | | | 0.078 | -0.102 | | | |
| | 8.1 | | | 0.0755 | -0.1 | | | |
| | 10 | | | 0.0765 | -0.102 | | | |
| | 12 | | | 0.0765 | -0.102 | | | |
| | 14 | | | 0.0742 | -0.102 | | | |
| | 16 | | | 0.078 | -0.107 | | | |
| | 18 | | | 0.0755 | -0.097 | | | |

| H (kOe) | T (K) | $\Gamma$ (meV) | $\Gamma_0$ (meV) | $\tilde{\lambda}_0$ | $\lambda_r$ | $\Delta$ (meV) | $\Gamma^D$ (meV) | Form of $N_S(E)$ |
|---|---|---|---|---|---|---|---|---|
| 20 | 0.45 | $10^4$ | 0.2 | 0.076 | -0.111 | 0.68 | 0.57 | $N_S^{BCS}(E)$ |
| | 0.9 | | | 0.076 | -0.111 | 0.47 | 0.63 | |
| | 1.5 | | | 0.076 | -0.111 | 0.43 | 0.63 | |
| | 2.0 | | | 0.076 | -0.111 | 0.36 | 0.58 | |
| | 2.5 | | | 0.076 | -0.111 | 0.35 | 0.8 | |
| | 2.8 | | | 0.0815 | -0.111 | | | Fitted with AA correction only |
| | 3.8 | | | 0.0772 | -0.111 | | | |
| | 4.5 | | | 0.0752 | -0.111 | | | |
| | 5.5 | | | 0.0757 | -0.111 | | | |
| | 6.0 | | | 0.0757 | -0.111 | | | |
| | 7.0 | | | 0.0762 | -0.111 | | | |
| | 7.5 | | | 0.0748 | -0.111 | | | |
| | 8.1 | | | 0.0745 | -0.111 | | | |

| H (kOe) | T (K) | $\Gamma$ (meV) | $\Gamma_0$ (meV) | $\tilde{\lambda}_0$ | $\lambda_r$ | $\Delta$ (meV) | $\Gamma^D$ (meV) | Form of $N_S(E)$ |
|---|---|---|---|---|---|---|---|---|
| 40 | 0.45 | $10^4$ | 0.2 | 0.073 | -0.111 | 0.58 | 0.79 | $N_S^{Mod}(E)$ |
| | 0.9 | | | 0.077 | -0.111 | 0.43 | 0.73 | |
| | 1.5 | | | 0.077 | -0.111 | 0.39 | 0.77 | |
| | 2.0 | | | 0.077 | -0.111 | 0.35 | 0.8 | |
| | 2.6 | | | 0.0805 | -0.111 | | | Fitted with AA correction only |
| | 3.8 | | | 0.0775 | -0.111 | | | |
| | 4.5 | | | 0.0775 | -0.111 | | | |
| | 5.5 | | | 0.0755 | -0.111 | | | |
| | 6.0 | | | 0.0765 | -0.111 | | | |
| | 6.5 | | | 0.0755 | -0.111 | | | |
| | 7.0 | | | 0.0755 | -0.111 | | | |

|  | 7.5 |  |  | 0.0765 | -0.111 |  |  |  |
|  | 8.1 |  |  | 0.0745 | -0.111 |  |  |  |

| H (kOe) | T (K) | $\Gamma$ (meV) | $\Gamma_0$ (meV) | $\tilde{\lambda}_0$ | $\lambda_r$ | $\Delta$ (meV) | $\Gamma^D$ (meV) | Form of $N_S(E)$ |
|---|---|---|---|---|---|---|---|---|
| 60 | 0.45 | $10^4$ | 0.2 | 0.072 | -0.111 | 0.54 | 1.35 | $N_S^{Mod}(E)$ |
|  | 0.9 |  |  | 0.084 | -0.111 |  |  |  |
|  | 1.5 |  |  | 0.0835 | -0.111 |  |  |  |
|  | 2.0 |  |  | 0.0825 | -0.111 |  |  |  |
|  | 2.5 |  |  | 0.0805 | -0.111 |  |  |  |
|  | 3.6 |  |  | 0.0785 | -0.111 | Fitted with AA correction only |  |  |
|  | 4.2 |  |  | 0.0785 | -0.111 |  |  |  |
|  | 5.2 |  |  | 0.0785 | -0.111 |  |  |  |
|  | 6.05 |  |  | 0.0785 | -0.111 |  |  |  |
|  | 7.0 |  |  | 0.0775 | -0.111 |  |  |  |
|  | 8.1 |  |  | 0.0765 | -0.111 |  |  |  |

| H (kOe) | T (K) | $\Gamma$ (meV) | $\Gamma_0$ (meV) | $\tilde{\lambda}_0$ | $\lambda_r$ | $\Delta$ (meV) | $\Gamma^D$ (meV) |
|---|---|---|---|---|---|---|---|
| 80 | 0.45 | $10^4$ | 0.2 | 0.08 | -0.111 | Fitted with AA correction only |  |
|  | 0.9 |  |  | 0.082 | -0.111 |  |  |
|  | 1.6 |  |  | 0.083 | -0.115 |  |  |
|  | 2.0 |  |  | 0.081 | -0.115 |  |  |
|  | 2.5 |  |  | 0.0805 | -0.11 |  |  |
|  | 3.76 |  |  | 0.0805 | -0.11 |  |  |
|  | 4.83 |  |  | 0.0805 | -0.11 |  |  |
|  | 6.05 |  |  | 0.079 | -0.11 |  |  |
|  | 8.1 |  |  | 0.08 | -0.111 |  |  |

| H (kOe) | T (K) | $\Gamma$ (meV) | $\Gamma_0$ (meV) | $\tilde{\lambda}_0$ | $\lambda_r$ | $\Delta$ (meV) | $\Gamma^D$ (meV) | Form of $N_S(E)$ |
|---|---|---|---|---|---|---|---|---|
| 0 | 0.45 | $10^4$ | 0.2 | 0.072 | -0.1 | 0.715 | 0.345 | $N_S^{BCS}(E)$ |
| 5 |  |  |  | 0.072 | -0.1 | 0.68 | 0.4 |  |
| 10 |  |  |  | 0.072 | -0.1 | 0.67 | 0.44 |  |
| 20 |  |  |  | 0.072 | -0.1 | 0.665 | 0.535 |  |
| 30 |  |  |  | 0.073 | -0.1 | 0.625 | 0.65 |  |
| 40 |  |  |  | 0.072 | -0.1 | 0.55 | 0.64 | $N_S^{Mod}(E)$ |
| 50 |  |  |  | 0.072 | -0.1 | 0.525 | 0.69 |  |
| 60 |  |  |  | 0.072 | -0.1 | 0.495 | 0.78 |  |
| 70 |  |  |  | 0.072 | -0.1 | 0.49 | 1.2 |  |
| 85 |  |  |  | 0.078 | -0.111 | Fitted with AA correction only |  |  |

## IV. Zero-point fluctuations of vortices

To realise the effect of quantum fluctuations on vortex dynamics, the fractional amplitude of quantum fluctuation of vortex $\Delta a$ (about its mean position) is calculated with respect to the vortex lattice constant, $a = 1.075\sqrt{\frac{\phi_0}{H}}$ ($\phi_0$ is the quantum flux and $H$ is the applied magnetic field). For this we assume that the vortex is confined in a cage formed by the repulsive interaction with other vortices. Details of this calculation is given in ref. [3]. By equating vortex oscillation energy $m_v \omega^2 (\Delta a)^2 = \hbar \omega$, one can show that,

$$\frac{\Delta a}{a} = \frac{\hbar^{1/2}}{(k_s m_v)^{1/4}} \frac{1}{\sqrt{a}}$$

Where $m_v$ is the mass of the vortex, $\omega$ is the oscillation frequency and $k_s = \frac{\phi_0^2 d}{16\pi \mu_0 \lambda^2}$. $d$ is the sample thickness and $\lambda$ is the zero-field penetration depth. For 2nm thick $a$-MoGe film $\lambda$=1162 nm[4]. On the other hand in ref. 3, the vortex mass in a 20 nm thick $a$-MoGe film was found to be $36 m_e$ where $m_e$ is the electron mass. Assuming that the vortex mass per unit length remains unchanged with decreasing thickness here $m_v \approx 3.6 m_e$. Using these parameters value, we obtain $\frac{\Delta a}{a} \approx 0.5$ at $H = H_c^*$.

## V. Scaling analysis of isothermal magnetoresistance

Since our results indicate that the S-BM transition corresponds to a quantum critical point, we attempt a scaling analysis of $R_s$-$H$ data close to $H_c^*$. The scaling behaviour can be expressed as[5],

$$R_s(H,T) = R_c^* f\left(C_a \frac{|H-H_c^*|}{T^{\frac{1}{zv}}}\right),$$

where, $R_c^*$ is the sheet resistance at the critical point and $zv$ is the product of dynamical critical exponent and correlation length exponent. We obtain $zv = 1.13$ from the slope of $\log\left(\frac{dR}{dH}\big|_{H=H_c^*}\right)$ vs $\log\left(\frac{1}{T}\right)$ using the relation[6], $\left.\frac{dR}{dH}\right|_{H=H_c^*} \propto R_c^* T^{-\frac{1}{\nu z}} f'(0)$ (inset Fig. S2). In Fig. S2 we plot $R_s$ as a function of $\frac{|H-H_c^*|}{T^{\frac{1}{z\nu}}}$ with

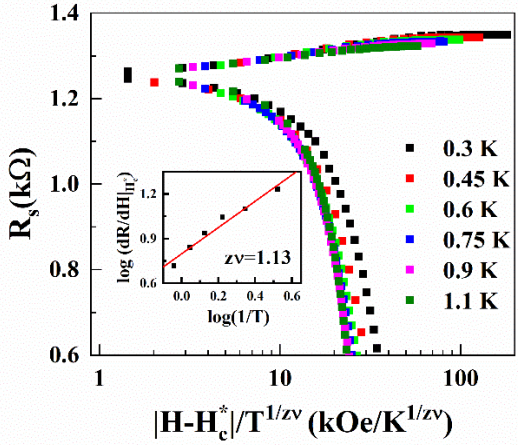

S2| plot of $R_s$ as function of scaling variable $\frac{|H-H_c^*|}{T^{1/z\nu}}$. *Inset*: linear fit of $\log\left(\frac{dR_s}{dH}|_{H_c}\right)$ vs $\log\left(\frac{1}{T}\right)$

this value of $z\nu$. The data collapse into two branches above and below in the vicinity of $H_c^*$; $z\nu$ is slightly lower than previously reported for MoGe films[7] of ~1.3, albeit for films with lower $T_c$.

[1] G. Blatter, M. V. Feigel'man, V. B. Geshkenbein, A. I. Larkin, and V. M. Vinokur, Vortices in high-temperature superconductors, Rev. Mod. Phys. **66**, 1125 (1994).

[2] B. Sacépé, C. Chapelier, T. Baturina, V. M. Vinokur, M. R. Baklanov and M. Sanquer Pseudogap in a thin film of a conventional superconductor. *Nat Commun* **1,** 140 (2010).

[3] S. Dutta, I. Roy, J. Jesudasan, S. Sachdev, and P. Raychaudhuri, *Evidence of zero-point fluctuation of vortices in a very weakly pinned a-MoGe thin film*, Phys. Rev. B **103**, 214512 (2021).

[4] S. Mandal, S. Dutta, S. Basistha, I. Roy, J. Jesudasan, V. Bagwe, L. Benfatto, A. Thamizhavel, and P. Raychaudhuri, *Destruction of superconductivity through phase fluctuations in ultrathin a-MoGe films*, Phys. Rev. B 102, 060501(R) (2020).

[5] M. P. A. Fisher, *Quantum phase transitions in disordered two-dimensional superconductors*, Phys. Rev. Lett. **65**, 923 (1990).

[6] A. F. Hebard and M. A. Paalanenn, *Magnetic-Field-Tuned Superconductor-Insulator Transition in Two-Dimensional Films*, Phys. Rev. Lett. **65,** 927 (1990).

[7] A. Yazdani and A. Kapitulnik, *Superconducting-Insulating Transition in Two-Dimensional a-MoGe Thin Films*, Phys. Rev. Lett. **74**, 3037 (1995).